\documentclass[prd,aps,10pt,twocolumn,nofootinbib,showpacs,notitlepage,amsmath,amssymb,aps]{revtex4-2}

\usepackage{hyperref}

\usepackage{graphicx}
\usepackage{dcolumn}
\usepackage{bm}
\usepackage{bbm}

\newcommand{\A}{\mathcal{A}}
\newcommand{\K}{\mathcal{K}}
\newcommand{\Hh}{\mathcal{H}}
\newcommand{\G}{\mathcal{G}}
\newcommand{\N}{\mathbb{N}}

\newcommand{\R}{\mathbb{R}}
\newcommand{\ie}{\textit{i}.\textit{e}.}
\newcommand{\Trb}[2]{\text{Tr}_{#1}\left[ #2 \right]} 

\newcommand{\ket}[1]{| #1 \rangle}
\newcommand{\bra}[1]{\langle #1 |}
\newcommand{\scalar}[2]{\langle #1 | #2 \rangle} 
\newcommand{\braket}[2]{\langle #1 | #2 \rangle}
\newcommand{\mean}[1]{\langle #1 \rangle}

\DeclareMathOperator{\Tr}{Tr}

\begin{document}

\title{Nonlocal correlations for semiclassical states in loop quantum gravity}	
\author{Filipe H. C. Menezes}
\email{fmenezes@fisica.ufmg.br}
\author{N. Yokomizo}
\email{yokomizo@fisica.ufmg.br}
\affiliation{
	Departamento de Física - ICEx, Universidade Federal de Minas Gerais\\CP 702, 30161-970, Belo Horizonte, MG, Brasil
	}
\date{\today}
	
\begin{abstract}
		We compute the two-point correlation function of the area operator for semiclassical states of loop quantum gravity in the limit of large spins. The cases of intrinsic and extrinsic coherent states are considered, along with a new class of semiclassical states constructed as perturbations of Livine-Speziale coherent states. For the usual coherent states, the correlations are shown to be short-ranged, decaying exponentially with the distance. Introducing perturbations given by correlated elementary excitations and decays of the gravitational field along pairs of loops, we obtain new states that, while preserving the peakedness properties of the unperturbed states, can also display long-ranged correlations. The perturbed coherent states include examples reproducing the typical decay of correlations for quantum fluctuations of the geometry associated with free gravitons on a background metric. Such a behavior is a natural requirement for the compatibility of semiclassical states in quantum gravity with the physical regime pictured by perturbative quantum gravity.
\end{abstract}
	
\maketitle

\section{Introduction}

Loop quantum gravity (LQG) is a theory of quantum gravity resulting from the quantization of general relativity described in terms of Ashtekar variables \cite{ashtekar2004background,rovelli2004quantum,thiemann2008modern,Perez:2012wv,Rovelli:2014ssa,Ashtekar:2021kfp}. As a necessary consistency requirement, the theory must reduce to general relativity in an adequate semiclassical regime. In its canonical formulation, the main applications of LQG include the description of quantum corrections to the dynamics of cosmological spacetimes \cite{Ashtekar:2021kfp,Li2024,Bojowald:2008zzb} and of black holes \cite{Ashtekar2023}, and in these cases of highly symmetric spacetimes, a consistent classical limit is observed, with the effective dynamics reducing to that of general relativity in the low-energy regime. In the covariant version of the theory described by the spinfoam formalism \cite{Perez:2012wv,Rovelli:2014ssa,Ashtekar:2021kfp}, the classical limit has also been established for simple cosmological models \cite{Bianchi:2010zs,Bianchi:2011ym}, and the EPRL vertex that encodes the dynamics of the theory was shown to reproduce the graviton propagator at large scales \cite{Bianchi:2006uf}, indicating its compatibility with semiclassical gravity. A complete description of the classical limit of LQG for generic spacetimes, however, is still an open question. A step towards this goal is the construction of adequate semiclassical states of the quantum geometry. In this work, we discuss the characterization of such states in the canonical formalism of LQG and introduce a new family of semiclassical states displaying a set of desired properties, including a condition on the decay of correlations in the fluctuations of the geometry.

In the canonical approach \cite{ashtekar2004background,rovelli2004quantum,thiemann2008modern}, constraints describing the classical equations of general relativity are promoted to operators, and physical states of the quantum geometry are required to satisfy these constraints. The kinematical Hilbert space $\K$ of LQG is the space of solutions to the Gauss and diffeomorphism constraints. These constraints select configurations of the spatial geometry at a given instant of time that are invariant under gauge transformations related to changes of local frames and under spatial diffeomorphisms. Physical states of the geometry must satisfy, in addition, the Hamiltonian constraint. The requirement of diffeomorphism invariance is a strong condition, and severely restricts the allowed representations of the algebra of observables of canonical gravity in Ashtekar variables given by the holonomy-flux algebra. In fact, according to a fundamental theorem presented in \cite{Lewandowski:2005jk}, the kinematics of LQG is uniquely fixed by the requirement of diffeomorphism and gauge invariance. This uniqueness theorem provides a well-motivated choice for the representation of the kinematics employed in both the canonical and covariant approaches in LQG. The quantization of the more intricate Hamiltonian constraint, on the other hand, involves ambiguities that are not yet well understood. As a result, the description of the space of solutions to the Hamiltonian constraint within the kinematical Hilbert space $\K$ is an open question in the theory, and the main missing step for the full implementation of the quantization program in the canonical setting (see review \cite{Ashtekar:2021kfp}).

A generic state of the quantum geometry in the kinematical Hilbert space $\K$ can be expanded in the spin-network basis. Operators describing the geometry, as areas, volumes and dihedral angles, have discrete spectra in the spin-network representation, and in general do not commute, leading to a representation of the spatial geometry in terms of superpositions of quantum discrete geometries. Among such quantum geometries, special classes of semiclassical states have been constructed that are peaked on classical configurations of the spatial geometry, with small fluctuations \cite{Thiemann:2000bw,Thiemann:2002vj,LSvertex,Freidel:2010tt,Coherentspinnetworks,Bonzom:2012bn}. Distinct families of semiclassical states are obtained depending on which properties of the geometry are required to display a semiclassical behavior. Intrinsic coherent states, which include Livine-Speziale coherent states \cite{LSvertex}, are peaked on classical configurations of the spatial intrinsic geometry. Extrinsic coherent states are peaked on both the spatial intrinsic and extrinsic geometry, the main examples of which being heat kernel states \cite{Thiemann:2000bw,Thiemann:2002vj,Coherentspinnetworks}. Definitions and properties of varied families of semiclassical states of LQG are discussed in \cite{Bonzom:2012bn,Bianchi:2016hmk}.

As the known semiclassical states of LQG are constructed in the kinematical Hilbert space, they do not necessarily satisfy the Hamiltonian constraint. The properties of peakedness and small fluctuations must naturally be required from semiclassical states, but it might be that additional properties are involved in the characterization of physical semiclassical states in the subspace of solutions of the Hamiltonian constraint within $\K$. In the absence of a  well-established form for the Hamiltonian constraint, such a question cannot be answered in definite form at this point, but guidance can be obtained from semiclassical gravity, described as an effective theory \cite{Donoghue:1994dn,Burgess:2003jk} in the context of quantum field theory on curved spacetimes \cite{Birrell:1982ix,Parker:2009uva}.

Considering perturbative quantum gravity, with gravitons propagating on a classical background spacetime, a key property of the quantum fluctuations of the gravitational field is that they are highly entangled, with correlations in the fluctuations of the field amplitude decaying polynomially at large distances as $1/d^2$. Such correlations are responsible for the emergence of an area law for the entanglement entropy \cite{Sorkin:1983,Bombelli:1986rw,Srednicki:1993im}, proposed as a possible source for black hole entropy \cite{Bombelli:1986rw}, and argued to provide a necessary criterion for the selection of semiclassical states in any theory of quantum gravity \cite{Bianchi:2012ev}. Indeed, states of the quantum geometry that do not display the typical decay of correlations with $1/d^2$ are incompatible with the semiclassical regime described by perturbative quantum gravity, the most reliable stepping stone towards full quantum gravity at the present. Accordingly, as proposed in \cite{Bianchi:2016tmw}, it is natural to search for semiclassical states in LQG that satisfy the additional property of displaying correlations decaying at large distances as in the continuum, beyond the usual requirements of peakedness and small fluctuations. Other strategies for the construction of semiclassical states consistent with the regime described by quantum field theory on curved spacetime were pursued in \cite{Chirco:2014naa,Sahlmann:2019elx}.

A large class of entangled states in LQG was introduced in \cite{Bianchi:2016tmw,Bianchi:2016hmk} by exploring the bosonic representation of LQG \cite{Girelli:2005ii,Freidel:2010tt,Borja:2010rc}. In this representation, states of the geometry are described in terms of excitations of a network of bosonic variables on the lattice. The analysis of correlations in bosonic lattices is a well-developed research area, and its techniques could be imported to the context of LQG. The simplest class of entangled states in bosonic lattices with a generic two-point correlation function consist of squeezed vacua \cite{Adesso:2014rev,Adesso:2007jg,Bianchi:2015fra,Bianchi:2017kgb}. Considering the analogous states in the bosonic representation of LQG and projecting them to the subspace selected by the kinematical constraints of the theory, a family of squeezed states of the geometry with tunable correlations was obtained in \cite{Bianchi:2016tmw}. In particular, squeezed vacuum states with correlations decaying as $\sim 1/d^2$ in the limit of small spins were found \cite{Bianchi:2016tmw}. On leaving the regime of small spins, however, the calculation of the correlations becomes prohibitively hard, due to the combination of a large number of excitations of elementary oscillators and the presence of constraints. Even in the limit of small spins, a large number of excitations is required to build simple gauge invariant excitations---eight oscillators must be excited in order to build a single loop state, for instance. Moreover, in order to reach a semiclassical regime of large spins, the number of excitations in a region must also be large.

In this work, we analyze the decay of correlations in fluctuations of the geometry for semiclassical states of LQG. We consider a regular cubic lattice and focus on the calculation of area-area correlations. Livine-Speziale coherent states and heat kernel states are analyzed as examples of intrinsic and extrinsic coherent states. We find that, in the limit of large spins, the area-area two-point correlation function decays exponentially on the lattice for both classes of states, with a correlation length of only a few sites, \ie, the correlations are short-ranged. A new family of entangled states is then introduced, and shown to display long-ranged correlations. In particular, states with an area-area two-point correlation function decaying as $1/d^2$ in the limit of large spins are identified. In contrast with the the squeezed vacuum states introduced in \cite{Bianchi:2016tmw}, obtained through the application of squeezing operators to the Ashtekar-Lewandowski vacuum state \cite{AL-vacuum,AL-vacuum-Baez}, here we consider states defined as perturbations of Livine-Speziale coherent states. In this way, the semiclassical geometry described by such states plays the role of a background geometry, over which correlated perturbations are introduced, in a picture reminiscent of perturbative quantum gravity. In addition, the perturbations are introduced so that the resulting states are automatically gauge invariant, which simplifies the analysis of the correlations. A state that is still peaked on a clasical geometry is obtained, but which displays long-ranged correlations that can be analytically computed in the limit of large spins.

This work is organized as follows. In Section \ref{sec:lqg}, we review basic elements of loop quantum gravity and the bosonic representation of its kinematical Hilbert space. In Section \ref{sec:coherent-states}, we compute area-area correlations for intrinsic and extrinsic coherent states on a cubulation. A new class of states constructed through a perturbation of Livine-Speziale coherent states is introduced in Section \ref{sec:correlated-coherent-states}, and shown to display long-ranged correlations in the fluctations of the geometry, while preserving the peakedness property of the unperturbed states. We conclude with a brief discussion of the results and future directions in Section \ref{sec:conclusion}.

\section{Loop quantum gravity and the bosonic representation}
\label{sec:lqg}

Loop quantum gravity is a model of quantum gravity obtained by the canonical quantization of general relativity formulated in terms of Ashtekar variables \cite{ashtekar2004background, rovelli2004quantum, thiemann2008modern}. In this work we are  interested in the analysis of correlations in quantum fluctuations of the geometry for semiclassical states. We start with a brief review of the elements of loop quantum gravity relevant for our purposes.

The kinematical Hilbert space $\K$ of loop quantum gravity is constructed from Hilbert spaces defined on oriented graphs. Let $\Gamma$ be an oriented graph with $N$ nodes $n$ and $L$ links $\ell$. We first define the Hilbert space $\Hh_\Gamma$ as:
\begin{equation}\label{hilbert_space_1}
    \Hh_\Gamma = \bigotimes_{\ell \in \Gamma} \bigoplus_{j_\ell} \Hh_{j_\ell} \otimes \Hh_{j_\ell},
\end{equation}
where $\Hh_{j_\ell}$ is the Hilbert space associated to the $SU(2)$ irreducible representation $j_\ell$, with $j_\ell$ different from zero. Each factor of $\Hh_{j_\ell}$ is associated with an endpoint of the link $\ell$. This Hilbert space admits an orthonormal basis $\{\ket{\Gamma, \{j_\ell, m_\ell, n_\ell\}}\}$ with basis states labeled by a spin $j_\ell$ and two magnetic numbers $m_\ell, n_\ell$ per link:
\begin{equation}\label{spinnetwork_non_invariant}
    \ket{\Gamma, \{j_\ell, m_\ell, n_\ell\}} := \bigotimes_{\ell\in\Gamma} \ket{j_\ell, m_\ell}\ket{j_\ell, n_\ell}.
\end{equation}

The next step in the construction of the kinematical Hilbert space $\K$ consists of defining invariant tensors, called intertwiners, at the nodes of the graphs. By reordering the spaces $\Hh_\ell$ on the right side of Eq.~\eqref{hilbert_space_1}, we can associate a Hilbert space $\Hh_n$ with each node $n \in \Gamma$:
\begin{equation}
    \Hh_n = \bigoplus_{j_\ell} \bigotimes_{\ell \in n} \Hh_{j_\ell},
\end{equation}
where $\ell \in n$ represents the links for which the node $n$ corresponds to an endpoint. Using Clebsch-Gordan relations, the Hilbert space $\Hh_n$ can be written as a direct sum of irreducible representations of $SU(2)$ over different spins $j$. An intertwiner $i_n$ is an invariant tensor, \ie, an element of the $SU(2)$-invariant subspace. For a given spin configuration $\{j_\ell\}$, the tensor product of intertwiners $i_n$ over all nodes of the graph produces the so-called spin-network states, denoted by
\begin{equation}	\label{eq:spin-network}
\ket{\Gamma, \{j_\ell, i_n\}} = \prod_n [i_n]^{m_{n1}, \dots, m_{n|m|}} \cdot \bigotimes_{\ell\in\Gamma} \ket{j_\ell, m_\ell}\ket{j_\ell, n_\ell} \, ,
\end{equation}
where the dot indicates that intertwiner indices must be contracted with magnetic numbers of the ket states according to the combinatorics of the graph. Such states form an orthogonal basis of the $SU(2)$-invariant subspace $\K_\Gamma \subset \Hh_\Gamma$, known as the kinematical Hilbert space on the graph $\Gamma$. A generic state $\ket{\psi} \in \K_\Gamma$ can be written as:
\begin{equation}
    \ket{\psi} = \sum_{j_\ell, i_n} c_{ \{j_\ell, i_n\} } \ket{\Gamma, \{j_\ell, i_n\} }.
\end{equation}
The kinematical Hilbert space of loop quantum gravity is the direct sum of $\K_\Gamma$ over all oriented graphs:
\begin{equation}	\label{eq:sum-over-Gamma}
    \K = \bigoplus_\Gamma \K_\Gamma.
\end{equation}
	
The kinematical Hilbert state can also be presented in the holonomy representation, $\K_\Gamma \simeq L^2[SU(2)^L/SU(2)^N]$, with states represented as square-integrable functions of holonomies $h_\ell$ living at the links of the graph,
\begin{equation}
\braket{h_\ell}{\psi} = \psi(h_\ell) \, ,
\end{equation}
which are invariant under gauge transformations
\begin{equation}	\label{eq:gauge-transf}
\psi(h_\ell) \mapsto (U_g \psi)(h_\ell) = \psi\left( U_{s(\ell)}h_\ell U_{t(\ell)}^{-1} \right) \, ,
\end{equation}
where $s(\ell)$ and $t(\ell)$ are the source and target nodes of the link $\ell$, and a local transformations $U_n \in SU(2)$ is associated with each node $n$ of the graph. The list $\{U_n, n \in \Gamma\}$, characterizes the gauge transformation $U_g$. A state in the spin-network representation is mapped to the holonomy representation through the unitary transformation \cite{Livine:2011gp,Livine:2011zz,Livine:2013zha}:
\begin{equation}	\label{eq:spin-h-map}
(-1)^{j_\ell-n_\ell} \ket{j_\ell,m_\ell} \ket{j_\ell,-n_\ell} 
\mapsto \sqrt{2j_\ell+1} [D^{j_\ell}(h_\ell)]^{m_\ell}{}_{n_\ell} \, ,
\end{equation}
where the $D^{j_\ell}(h_\ell)$ are Wigner matrices.

\subsection{Bosonic representation}

States in the kinematical Hilbert space $\mathcal{K}$ admit an alternative representation in terms of bosonic variables in the so-called bosonic representation of loop quantum gravity \cite{Girelli:2005ii,Freidel:2010tt,Borja:2010rc}. This formalism is an adaptation of the Schwinger oscillator model of angular momentum \cite{Schwinger:1952dse,Sakurai:2011zz} to the context of loop quantum gravity. Alternatively, it can be seen as the canonical quantization of the classical phase space of LQG described in terms of spinorial variables \cite{Livine:2011gp,Livine:2011zz,Livine:2013zha}. For a review, see \cite{Bianchi:2016hmk}.

Let us describe the bosonic representation for the kinematical Hilbert space $\mathcal{K}_\Gamma$ on a fixed graph $\Gamma$. The full kinematical Hilbert space $\mathcal{K}$ is then obtained by taking the direct sum over all graphs, as specified in Eq.~\eqref{eq:sum-over-Gamma}. We first introduce a set $\mathcal{S}$ of $2L$ elements, called seeds and labeled by $i = 1, ...,2L$. Each seed corresponds to an endpoint of a link. We say that $i \in n$ if the endpoint associated with $i$ is the node $n$. The valence of the node is the number of links meeting at it, which we denote by $|n|$. Ordering the seeds at a node with an index $\mu=1,\dots,|n|$, the seeds can be alternatively labeled by a pair $n\mu$. Each link $\ell$ is associated with a pair of seeds, that we label by source $s(\ell)$ and target $t(\ell)$, according to the link orientation. 
If two seeds, $i$ and $j$, are associated with the same node, we say that $\left<i,j\right>$ is a wedge.
	
With each seed, we associate a pair of creation and annihilation operators, labeled by an index $A=0,1$, which satisfy the canonical commutation relations
\begin{equation}
[a_i^A, a_j^{B\dagger}] = \delta_{ij}\delta^{AB}, \quad [a_i^A, a_j^B] = 0, \quad [a_i^{A\dagger}, a_j^{B\dagger}] = 0.
\end{equation}
We define the bosonic Hilbert space $\Hh_{\mathcal{S}}$ as the bosonic Fock space built over the global vacuum $\ket{0}$, defined as
\begin{equation}
		a^A_i \ket{0} = 0, \quad i=1,...,2L, \quad A = 0, 1.
\end{equation}
The global vacuum is the tensor product of the vacua of all local bosonic variables. The bosonic Hilbert space is the tensor product of local bosonic spaces at each seed:
\begin{equation}	\label{eq:bosonic-space}
\mathcal{H}_\mathcal{S} = \bigotimes_{i=1}^{2L} \mathcal{H}_i 
, ,
\end{equation}
where each $\mathcal{H}_i$ is the Fock space associated with the bosonic variables $a_i^A$, $A=0,1$. The space naturally factors over the nodes of the graph:
\begin{equation}	
\mathcal{H}_\mathcal{S} = \bigotimes_{n=1}^{N} \mathcal{H}_{\mathcal{S}, n} \, , \quad  \mathcal{H}_{\mathcal{S}, n} = \bigotimes_{\mu=1}^{|n|} \mathcal{H}_{n \mu} \, .
\end{equation}

At each seed, a unitary representation of $SU(2)$ can be constructed from the bosonic creation and annihilation operators. The generators $\vec{J}_i$ and the spin operator $\mathcal{J}_i$ are given by
	\begin{equation}
		\vec{J}_i := \frac{1}{2} \vec{\sigma}_{AB} a_i^{A\dagger} a_i^{B}, \qquad \mathcal{J}_i := \frac{1}{2} \delta_{AB} a_i^{A\dagger} a_i^{B}, 
	\end{equation} 
where $\vec{\sigma}$ are the Pauli matrices, and the indices $A$, $B$ are raised, lowered and contracted with the Kronecker delta. The operators $\vec{J}_i$ so constructed satisfy the usual commutation relations
	\begin{equation}
		[J^a_i, J^b_i] = i {\epsilon^{ab}}_c J^c_i
	\end{equation}
of the Lie algebra $su(2)$. The Casimir operator
\begin{equation}
J_i = \sqrt{\vec{J}_i \cdot \vec{J}_i}
\end{equation}
is written in terms of the spin operator as
\begin{equation}
J_i = \sqrt{\mathcal{J}_i (\mathcal{J}_i + 1)}
\end{equation}

The holonomy operator $h_\ell$ is defined at each link $\ell=(s,t)$ as:
\begin{align}
	{(h_\ell)^A}_B &= (2 \mathcal{J}_t+1)^{-\frac{1}{2}} \nonumber \\
		& \quad \times \left( \epsilon^{AC} a_{tC}^\dagger a_{sB}^\dagger - \epsilon_{BC} a_{t}^A a^C_s \right) (2 \mathcal{J}_s+1)^{-\frac{1}{2}} \, .
\end{align}
Under the condition $\mathcal{J}_{s(\ell)} = \mathcal{J}_{t(\ell)}$, the holonomy operator and the generators $\vec{J}_i$ satisfy the commutation relations
	\begin{align}
		[\vec{J}_s, h_\ell] &= \frac{1}{2} h_\ell \vec{\sigma}, \nonumber \\
		[\vec{J}_t, h_\ell] &= -\frac{1}{2} \vec{\sigma} h_\ell, \nonumber \\
		[{(h_\ell)^A}_B, {(h_{\ell'})^C}_D] &= 0 \, ,
	\end{align}
which correspond to the holonomy-flux algebra of observables of loop quantum gravity. In addition, a spin operator associated with individual links can then be introduced:
\begin{equation}
\mathcal{J}_\ell = \mathcal{J}_{s(\ell)}=\mathcal{J}_{t(\ell)} \, ,
\end{equation}
as well as Casimirs associated with links:
\begin{equation}
J_\ell = J_{s(\ell)}=J_{t(\ell)} \,.
\end{equation}
The area operator $\A_\ell$ for a single link $\ell$ is proportional to the Casimir operator $J_\ell$ \cite{ashtekar2004background,rovelli2004quantum, thiemann2008modern,Ashtekar:1996eg},
\begin{equation}	\label{eq:area-op}
\A_\ell = a_0 J_\ell \, , 
\end{equation}
where $a_0 = 8 \pi G \hbar \gamma_0$ is the area gap expressed in terms of the Barbero-Immirzi parameter $\gamma_0$.

The bosonic states are associated with spin states in the magnetic number basis through the Schwinger map:
\begin{equation}
\ket{j_i,m_i} =  \frac{(a_i^{0\dagger})^{j_i-m_i}}{\sqrt{(j_i-m_i)!}}\frac{(a_i^{1\dagger})^{j_i+m_i}}{\sqrt{(j_i+m_i)!}} \, \ket{0} \, .
\end{equation}
Under this identification, we can write:
\begin{equation} \label{eq:seed-space}
\qquad \Hh_i \simeq \bigoplus_{j_i} \mathcal{H}_{j_i} \, .
\end{equation}
The action of the Casimir and spin operators on the basis vectors reads:
\begin{align}
J_i \ket{j_i,m_i} &= \sqrt{j_i (j_i+1)} \ket{j_i,m_i} \, ,  \nonumber \\
\mathcal{J}_i \ket{j_i,m_i} &= j_i \ket{j_i,m_i} \, .
\end{align}

Imposing the restriction $\mathcal{J}_{s(\ell)} = \mathcal{J}_{t(\ell)}$ and $SU(2)$ invariance, we obtain an alternative representation of the kinematical Hilbert space $\K_\Gamma$ of loop quantum gravity on a graph $\Gamma$ as the subspace 
\begin{equation}	\label{eq:kinematical-bosonic}
\K_\Gamma = P_A P_G \, \Hh_{\mathcal{S}}
\end{equation}
satisfying the constraints:
\begin{equation}\label{vínculos}
		C_\ell := \mathcal{J}_{s(\ell) - \mathcal{J}_{t(\ell)}} \approx 0, \quad \vec{G}_n := \sum_{i \in n} \vec{J}_i \approx 0 \, 
\end{equation}
where $P_A$ is the projector to the subspace of spaces satisfying the area matching constraints $C_\ell$, and $P_G$ is the projector to the subspace of states satisying the Gauss constraints $\vec{G}_n$. The area matching constraint $C_\ell$ imposes the matching of the $SU(2)$ representations at the source and target seeds of the same link $\ell$, and the Gauss constraint $\vec{G}_n$ imposes the invariance under $SU(2)$ gauge transformations at the nodes $n$, so that the states are described by intertwiners at each node. It is convenient to introduce the projector $P_n: \mathcal{H}_{\mathcal{S}, n} \to \mathcal{H}_{\mathcal{S},n}$ to the intertwiner space at each node, which allows us to write:
\[
P_G = \bigotimes_{n=1}^N P_n \, .
\]

The operators $a_i^A$, $a_i^{A\dagger}$ can also be used to define creation and annihilation operators associated with wedges:
	\begin{equation}
		F_{ij} = \epsilon_{AB} a_i^A a_j ^B, \quad F_{ij}^\dagger = \epsilon_{AB} a_i^{A\dagger} a_j^{B\dagger}
	\end{equation}
where $\epsilon_{AB}$ is the Levi-Civita symbol with $\epsilon_{01} = 1$. If $i, j$ are seeds of distinct nodes, we set $F_{ij} = 0$. In addition, if $\alpha$ is an oriented loop, \ie, a set of oriented wedges, $\alpha = \{ w_1, ..., w_{|\alpha|} \}$ such that consecutive wedges ($w_i, w_{i+1}$ or $w_{|\alpha|},w_1$) are in adjacent nodes, we define annihilation and creation operators associated with $\alpha$:
	\begin{equation}
		F_\alpha = \prod_{i=1}^{|\alpha|} F_{w_i}, \quad F_\alpha^\dagger = \prod_{i=1}^{|\alpha|} F_{w_i}^\dagger,
	\end{equation}
The operator $F_\alpha^\dagger$ creates loop states
	\begin{equation}\label{loop states}
		\ket{\alpha} \propto F_\alpha^\dagger \ket{0}
	\end{equation}
that obey the Gauss and link constraints. Two-loop states are defined as
\begin{equation}
\ket{\alpha_1 \alpha_2} \propto F^\dagger_{\alpha_1}F^\dagger_{\alpha_2} \ket{0} \, .
\end{equation}
Generic multiloop states with any number of loop excitations are similarly defined, and can be used to construct an overcomplete basis of the kinematical Hilbert space $\K_\Gamma$ \cite{Bianchi:2016hmk}. In the holonomy representation, a loop state $\ket{\alpha}$ is the trace of the product of holonomies along the loop. If the links of the loop $\alpha$ are ordered as $\ell_{\alpha 1},\dots, \ell_{\alpha 4}$, we have:
\begin{equation}	\label{eq:one-loop-h-rep}
\psi_\alpha(h_\ell) :=\braket{h_\ell}{\alpha} = \Tr{h_\alpha} \, ,
\end{equation}
where $h_\alpha = h_{\ell_{\alpha 1}} h_{\ell_{\alpha 2}} h_{\ell_{\alpha 3}} h_{\ell_{\alpha 4}}$, and
\begin{equation}	\label{eq:two-loop-h-rep}
\psi_{\alpha_1, \alpha_2}(h_\ell)= \psi_{\alpha_1}(h_\ell) \psi_{\alpha_2}(h_\ell) \, .
\end{equation}

\section{Area-area correlations for coherent states}
\label{sec:coherent-states}
	
Coherent spin network states were discussed in several works \cite{Thiemann:2000bw,Thiemann:2002vj,LSvertex,Freidel:2010tt,Coherentspinnetworks,Bonzom:2012bn,Bianchi:2016hmk}. Distinct families of coherent states have been considered, depending on which properties of the geometry are required to be semiclassical. Intrinsic coherent states obtained by gluing together local Livine-Speziale intertwiners \cite{LSvertex} are semiclassical with respect to the intrinsic geometry. They have probability distributions that are sharply peaked at classical configurations of the intrinsic geometry, but display large fluctuations for the extrinsic geometry. We will call them Livine-Speziale (LS) coherent states. Extrinsic coherent states are semiclassical with respect to both the intrinsic and extrinsic geometries \cite{Thiemann:2000bw,Thiemann:2002vj,Coherentspinnetworks}. They are often called heat kernel states, and have been extensively studied \cite{Thiemann:2000ca,Thiemann:2000bx,Thiemann:2000by,Bahr:2007xa,Bahr_2009}. A review of the commonly used coherent spin network states can be found in \cite{Bonzom:2012bn,Bianchi:2016hmk}, to which we refer for details.

We are interested in analyzing area-area correlations for fluctuations of the spatial geometry in coherent spin network states on a fixed lattice $\Gamma$ in the limit of large areas with respect to the Planck scale.  Two specific families of coherent states will be considered: LS coherent states \cite{LSvertex} and heat kernel states \cite{Coherentspinnetworks,Thiemann:2000bw}. We will compute the large-distance behavior of the area-area correlations for these semiclassical states, exploring asymptotic approximations previously employed for the calculation of average values in the limit of large spins. It is convenient to choose a regular lattice $\Gamma$ for this purpose. We let $\Gamma$ be a graph with $N^3$ points dual to a cubic lattice. A well-behaved limit of $N \to \infty$ is taken at the end of the calculation. We assume periodic boundary conditions, so that points in opposite boundaries of the graph are connected by a link. Each node $n$ corresponds to a cube in the lattice, and each of the six links $n\mu$, $\mu=1,\ldots,6$, emanating from the node $n$ corresponds to a face in the cube. Since each face is shared by two cubes, there are two distinct labels $n\mu$ associated with any link, that is, $n\mu$ labels seeds.

\subsubsection{LS coherent states}

An LS coherent spin network state is uniquely determined by a set of parameters $\{(\lambda_{n\mu},\vec{v}_{n\mu})\}$. The unit vector $\vec{v}_{n\mu}\in\R^{3}$ represents the direction normal to the face dual to the link $n\mu$, as seen from the node $n$, and the real parameters $\lambda_{n\mu}\in\R$ are related to the average value of the areas. The normal directions can be represented by a set of spinors $\ket{z_{n\mu}}$ such that:
\begin{equation}	\label{eq:n-z}
\vec{\sigma}\cdot\vec{v}_{n\mu}\ket{z_{n\mu}}=\ket{z_{n\mu}} \, ,
\end{equation}
where $\vec{\sigma}$ is the vector formed by the Pauli matrices.
The spinors $\ket{z_{n\mu}}$ are written in the basis of eigenstates of the $\sigma_3$ operator as:
\begin{equation}
\ket{z_{n\mu}}=z_{n\mu}^0\ket{+}+z_{n\mu}^1\ket{-}, \quad |z_{n\mu}^0|^2+|z_{n\mu}^1|^2=1 \, .
\end{equation}

A bosonic coherent state $\ket{\lambda_{n\mu},\vec{v}_{n\mu}}$ is first associated with each seed:
\begin{equation} \label{eq:coherent-seed}
\ket{\lambda_{n\mu},\vec{v}_{n\mu}} = \exp \left[\lambda_{n\mu}z_{n\mu}^A\left(a_{n\mu}^A\right)^\dagger\right] \ket{0}_{n\mu} \, .
\end{equation}
Taking the tensor product of the bosonic coherent states over all seeds, area matching at each link, and group averaging at every node, an LS coherent spin network state is then defined as:
\begin{equation}\label{eq:def-coherent-states}
\ket{LS,\{\lambda_{n\mu},\vec{v}_{n\mu}\}} = P_{G} P_{A }\bigotimes_{n\mu}\ket{\lambda_{n\mu},\vec{v}_{n\mu}} \in \mathcal{K}_\Gamma \, ,
\end{equation}
where $P_{A}$ is the area matching projector and $P_{G}$ is the Gauss projector.

If the spins are fixed at all links, and consequently at all seeds, $j_{s(\ell)}=j_{t(\ell)}=j_\ell$, we obtain an LS coherent state at fixed spins,
\begin{align}	\label{eq:LS-fixed-spins}
P_{ \{j_\ell\}} \ket{LS,\{\lambda_{n\mu},\vec{v}_{n\mu}\}} &\propto \bigotimes_{n\mu} \ket{ LS,\{j_{n \mu}, \vec v_{n \mu} \} }_n \, ,
\end{align}
which, up to normalization, is a tensor product of LS coherent intertwiners
\begin{equation}
\ket{ LS,\{j_{n \mu}, \vec v_{n \mu} \} }_n = P_n  \bigotimes_{\mu=1}^{|n|} \ket{j_{n\mu},z_{n \mu}}_{n \mu} \, ,
\end{equation}
where
\begin{equation}
\ket{j,z} = \frac{1}{\sqrt{(2j)!}} \left( z^A a_A^\dagger \right)^{2j} \ket{0}
\end{equation}
is a normalized $SU(2)$ coherent state of spin $j$.

\subsubsection{Heat kernel states}

A heat kernel state $\ket{\{H_\ell,t_\ell\}}$ is labeled by a set of elements $H_\ell$ of the group $SL(2,\mathbb{C})$ and real parameters $t_\ell \in \mathbb{R}$ attached to the links of the graph. A generic element $H_\ell$ of $SL(2,\mathbb{C})$ can be parametrized as \cite{Coherentspinnetworks}:
\begin{equation}	\label{eq:sl2c-decomp}
H_\ell = R_{\vec{v}_s} e^{-iz_\ell \sigma_3/2} R_{\vec{v}_t}^{-1} \, , \quad z_\ell = \xi_\ell + i \eta_\ell \, ,
\end{equation}
where $R_{\vec{v}}$ is an element of $SU(2)$ that rotates a unit vector parallel to the axis $z$ into a unit vector $\vec{v}$, $\xi_\ell$ is an angle and $\eta_\ell$ is a real parameter. A local link state is first defined at each link $\ell$ as:
\begin{multline}	\label{eq:hk-def-1}
\ket{H_\ell,t_\ell} = \sum_{j_\ell,m_\ell,n_\ell} (2j_\ell+1) e^{-t_\ell j_\ell(j_\ell+1)} \\ \
	\times \left[D^{(j_\ell)}(H_\ell)\right]_{m_\ell n_\ell} \ket{j_\ell,m_\ell}\ket{j_\ell,n_\ell}\, ,
\end{multline}
where $D^{(j_\ell)}$ is the extension of the irreducible representation of spin $j_\ell$ of $SU(2)$ to its complexification $SL(2,C)$, with the Pauli matrices given as usual, and a generic group element of the form \eqref{eq:sl2c-decomp}. A heat kernel state on the graph is defined as the projection of the tensor product of such link states into $\mathcal{K}_\Gamma$:
\begin{equation}\label{eq:hk-def-2}
\ket{HK,\{H_\ell,t_\ell\}} = P_{G} P_{A }\bigotimes_{\ell} \ket{H_\ell,t_\ell} \, .
\end{equation}

\subsection{Homogeneous background}
\label{sec:homogeneous case}

In this section, we construct coherent states peaked at discrete classical geometries describing cubic lattices with periodic boundary conditions. The average geometry of such a state provides a representation of a homogeneous background, over which we will later analyze correlations for fluctuations of the geometry. We first consider LS coherent states and then heat kernel states.

LS coherent states are labeled by parameters $\{(\lambda_{n\mu},\vec{v}_{n\mu})\}$. We choose, at each node,  normals parallel to the Euclidean axes:
\begin{align}	\label{eq:cubic-pars-1}
\vec{v}_{n1}=-\vec{v}_{n4}=\frac{\hat{x}}{2} \, , \nonumber \\
\vec{v}_{n2}=-\vec{v}_{n5}=\frac{\hat{y}}{2}\, , \nonumber \\
\vec{v}_{n3}=-\vec{v}_{n6}=\frac{\hat{z}}{2} \, .
\end{align}
In addition, we set:
\begin{equation}	\label{eq:cubic-pars-2}
\lambda _{n\mu }= \lambda \, , \quad \forall n, \mu \, .
\end{equation}
The spinors associated to this choice of normal directions through Eq,~\eqref{eq:n-z} are denoted by $z_\mu$, since they are independent of the node $n$, $\vec{v}_{n\mu}=\vec{v}_\mu$. The factors of $1/2$ in the normals ensure that the corresponding spinors are normalized, $\langle z_\mu|z_\mu\rangle=1$. The parameter $\lambda$ will determine the average area of the faces. The normals are chosen parallel to the Euclidean axes so that a lattice with cubic geometry is obtained.

Let us first discuss the properties of the link states before the projection to the gauge-invariant sector implemented by the Gauss constraint $P_G$. This will prove to be useful for the determination of the probability distribution of spins for the LS coherent states at large quantum numbers. At a link $\ell$, consider the area matched tensor product of local bosonic coherent states \eqref{eq:coherent-seed} at its two endpoints. From Eq.~\eqref{eq:cubic-pars-2}, the parameters $\lambda$ at the source and target seed of the link are the same, and the link state has the generic form:
\begin{multline}\label{eq:link-state}
		P_A \left(\ket{\lambda_{s(\ell)},\vec{v}_{s(\ell) }} \otimes \ket{\lambda_{t(\ell)},\vec{v}_{t(\ell)}}\right) \\
		= \sum_{j\in \frac{\N}{2}} \frac{\lambda^{4j}}{(2j)!}\ket{j,z_{s(\ell)}} \ket{j,z_{t(\ell)}} \, .
\end{multline} 
The probability distribution for the spins is given by:
\begin{equation}
P(j)=\frac{\lambda^{8j}}{[(2j)!]^2}\frac{1}{I_0(2\lambda^2)} \, ,
\end{equation}
where $I_n$ is a modified Bessel function of the first kind, which leads to:
\begin{equation}
\mean{j} = \frac{\lambda^2}{2}\frac{I_1(2\lambda^2)}{I_0(2\lambda^2)}, \quad \mean{j^2} = \frac{\lambda^4}{4} \, .
\end{equation}

For large $\lambda \gg 1$ \cite{abramowitz},
\begin{equation}
I_0(2 \lambda^2) \simeq \frac{e^{2 \lambda^2}}{\sqrt{4 \pi \lambda^2}}
\end{equation}
and we obtain
\begin{equation} 
P(j) \simeq \frac{\lambda^{8j}}{[(2j)!]^2}e^{-2 \lambda^2} \sqrt{4 \pi \lambda^2} \, ,
\end{equation}
which is proportional to a squared Poisson distribution. But for large mean values, the Poisson distribution is well approximated by a Gaussian distribution:
\begin{equation}
\frac{e^{-\mu} \mu^x}{x!} \simeq \frac{1}{\sqrt{2 \pi \mu}} \exp\left[ -\frac{1}{2} \left( \frac{x-\mu}{\sqrt{\mu}} \right)^2  \right] \, ,
\end{equation}
and it follows that the probability distribution for the spins reduces to a Gaussian,
\begin{equation}\label{eq:link-gaussian}
P(j) \simeq \frac{1}{\sqrt{\pi}\lambda} \exp \left[-\frac{4}{\lambda^2}\left(j-\frac{\lambda^2}{2}\right)^2\right]
\end{equation}
with
\begin{equation}
\mean{j} \simeq \frac{\lambda^2}{2} =: j_0, \quad \sigma_j \simeq \frac{\lambda}{2\sqrt{2}} = \frac{\sqrt{j_0}}{2} \, .
\end{equation}

Let us now analyze the quantum geometry at the individual nodes, after the projection to the gauge-invariant sector through the application of the Gauss projector $P_G$, but before the application of the area-matching projector $P_A$, as another step towards the determination of the spin probability distribution at large quantum numbers.

For a fixed set of spins, the state at the node $n$ is a coherent $LS$ intertwiner:
\begin{equation}\label{eq:LS-node}
	\ket{ LS,\{j_{n \mu}, \vec v_{n \mu} = \vec v_\mu\} }_n = P_n\bigotimes_{\mu=1}^6 \ket{j_{n\mu},z_\mu} \, .
\end{equation}
When all spins are large, the squared norm of the LS intertwiner is well approximated by \cite{LSvertex}:
\begin{align*}
f(j_{n \mu}) 	:=& _n\scalar{LS,\{j_{n \mu}, \vec v_{n \mu}\} }{LS,\{j_{n \mu}, \vec v_{n \mu}\} }_n \\
		\simeq& \frac{1}{\sqrt{\pi \det H}} \exp(- H^{-1}_{ab} N^a N^b ) \, ,
\end{align*}
where $\vec N = \sum_\mu j_{n \mu} \vec v_{n \mu}$, and the Hessian $H_{ab}$ is defined as:
\[
H_{ab} = \sum_\mu j_{n \mu} P_{ab}^\mu \, , \quad P_{ab}^\mu = \delta_{ab} - v_{n \mu}^a v_{n \mu}^b \, .
\]
In our case, we find for the Hessian:
\begin{align*}
H_{ab} &= \textrm{diag}(J_n-J_{n1},J_n-J_{n2},J_n-J_{n3}) \, , \\
J_{na} &= j_{n a} + j_{n,a+3} \, , \quad J_n=\sum_\mu j_{n \mu} \, ,
\end{align*}
which leads to:
\begin{multline}
f(j_{n \mu}) \simeq \frac{1}{\sqrt{\pi}} \frac{1}{\sqrt{(J_n-J_{n1})(J_n-J_{n2})(J_n-J_{n3})}} \\
	\times \exp\left[ -\frac{(j_{n 1}-j_{n 4})^2}{J_n-J_{n1}} - \frac{(j_{n 2}-j_{n 5})^2}{J_n-J_{n2}} - \frac{(j_{n 3}-j_{n 6})^2}{J_n-J_{n3}}\right] \, .
\label{eq:ls-gaussian}
\end{multline}

The formula for the squared norm of the intertwiners can be further simplified by noticing that when the links attached to the intertwiner are coherent states of the form \eqref{eq:link-state} with a large parameter $\lambda$, Eq.~\eqref{eq:link-gaussian} implies that the relevant spin configurations are such that:
\[
j_{n \mu} = j_0 + d_{n \mu} \, , \quad \text{with } d_{n \mu} \sim \sqrt{j_0} \ll j_0 = \lambda^2/2 \, .
\]
In this regime, Eq.~\eqref{eq:ls-gaussian} reduces to:
\begin{align}
f(j_{n \mu}) &\simeq \frac{1}{\sqrt{\pi} (2 \lambda^2)^{3/2}} \, e^{-N^2/2\lambda^2} \nonumber \\
	&= \frac{1}{\sqrt{\pi} (2 \lambda^2)^{3/2}} \, \prod_a e^{-\Delta_{na}^2/2\lambda^2} \, ,
\label{eq:ls-gaussians}
\end{align}
with
\begin{equation}	\label{eq:delta-def}
\Delta_{na}=j_{na}-j_{n,a+3} \, .
\end{equation}
The factorization of the exponential in this formula means that the spin fluctuations in distinct axes are independent. From Eqs.~\eqref{eq:link-gaussian} and \eqref{eq:ls-gaussians}, we obtain for the probability associated with a given coloring $\{j_\ell\}$ of the links:
\begin{multline}	\label{eq:spin-prob-coherent}
P_\lambda(\{j_\ell\}) \propto \prod_{\text{links}} \exp \left[ - \frac{4}{\lambda^2} \left(j_\ell - \frac{\lambda^2}{2}  \right)^2 \right] \\
\times \prod_{\text{nodes}} \left( \prod_a e^{-\Delta_{na}^2/2\lambda^2}\right).
\end{multline}
The probability distribution is a Gaussian function in the space of colorings $\{j_\ell\}$, peaked at the average values $\mean{j_\ell} =j_0=\lambda^2/2$. The quadratic expression in the exponent of the Gaussian does not involve products of spins at links pointing in nonparallel directions (such as $j_{m1} j_{n5}$, for instance).

Let us now construct heat kernel states with intrinsic geometries peaked on cubic lattices. A heat kernel state is specified by a set of parameters $\{ \vec{v}_{s(\ell)}, \vec{v}_{t(\ell)},\xi_\ell, \eta_\ell,t_\ell \}$, as described in Eqs.~\eqref{eq:sl2c-decomp}--\eqref{eq:hk-def-2}. We choose the normals $\vec{v}_{s(\ell)}, \vec{v}_{t(\ell)}$ as done for the LS coherent states, so that these directions represent again normals to the faces of a cubic lattice. The parameters $\xi_\ell, \eta_\ell,t_\ell$ are chosen independently of $\ell$:
\begin{equation}	\label{eq:hs-pars-2}
\xi_\ell=\xi, \quad \eta_\ell=\eta, \quad t_\ell=t \, .
\end{equation}
Introducing the parameter $j_0$ through:
\[
2 j_0 + 1 = \frac{\eta}{t} \, ,
\]
the asymptotic form of the heat kernel states for configurations peaked at large spins, obtained for $j_0 \gg 1$, is given by \cite{Coherentspinnetworks}:
\begin{multline}	\label{eq:HT-prob-gaussian}
\ket{\{H_\ell,t_\ell\}} \propto \sum_{\{j_\ell\}} \left( \prod_\ell (2j_\ell +1) e^{-t (j_\ell - j_0)^2} e^{- i \xi j_\ell} \right) \\
	\times \bigotimes_{i=1}^{N} \ket{ LS,\{j_{n \mu}, \vec v_{n \mu}\} }_n \, .
\end{multline}
The probability distribution for the spin configurations is:
\begin{equation}
P(\{j_\ell\}) \propto \prod_{\text{links}} \exp \left[ - 2 t (j_\ell -j_0)^2\right] \prod_{\text{nodes}} \left( \prod_a e^{-\Delta_{na}^2/4j_0}  \right) ,
\end{equation}
where we have used the asymptotic formula for the LS coherent intertwiners, fixed $\vec v_{n \mu}= \vec v_\mu$, and set $j_\ell \simeq j_0$ except in the exponents of the Gaussian functions.

\subsection{Decay of correlations}
\label{sec:correlations decay}

In the regime of large spins, $j_\ell \gg 1$, both LS coherent states and heat kernel states have Gaussian probability distributions for the spins. Exploring this asymptotic regime of large quantum numbers, we now proceed to the calculation of area-area correlations, considering the case when the states are peaked on a cubic lattice, which we discussed in the previous section, and focusing on the large-distance behavior of the correlations. We are interested in determining how the correlations decay with respect to the distance between the considered faces, dual to links of the graph.

From Eq.~\eqref{eq:area-op}, at each link, the area operator $\mathcal{A}_\ell$ is proportional to the Casimir operator $J_\ell$, with a proportionality constant given by the area gap $a_0 = 8 \pi G \hbar \gamma_0$. The correlation function of the area operator is thus given by:
\begin{equation}\label{eq:area-correlation-function}
		\G(\A_\ell, \A_{\ell'}) = a_0^2 \, \G(J_\ell, J_{\ell'}) \, ,
\end{equation}
where
\begin{equation}\label{eq:correlation-function}
		\G(J_\ell, J_{\ell'}) = \mean{J_\ell J_{\ell'}} - \mean{J_{\ell}} \mean{J_{\ell'}} \, .
\end{equation}
In the limit of large spins, $J_\ell \simeq \mathcal{J}_\ell$, and the correlation function of the Casimir operator $J_\ell$ is well approximated by that of the spin operator $\mathcal{J}_\ell$.

On a cubic lattice, a natural notion of distance between links is available. For concreteness, embed the cubic lattice in $\mathbb{R}^3$ by representing its nodes as points $n= (n_1,n_2,n_3)$, with $n_1,n_2,n_3= 1, \ldots, L$. If $\ell=n \mu$ and $\ell'=n' \mu'$ are links dual to the faces of interest, the lattice distance between them is defined as
\begin{equation}
d_{\ell \ell'}(\lambda) = \sqrt{\mean{\A}} \sqrt{(n_1-n'_1)^2+(n_2-n'_2)^2+(n_3-n'_3)^2} \, ,
\end{equation}
where $\mean{\A}= \mean{\A_\ell}= a_0 \mean{J_\ell}$ is the mean area at a link $\ell$, which is independent of the link in a homogeneous geometry. Therefore, by computing the averages $\mean{J_\ell}$ and correlation function $\G(J_\ell, J_{\ell'})$ of the spin operator, we can determine how the correlations in area fluctuations decay with the distance on the lattice. Moreover, in the limit of large spins, which is our main interest, these calculations reduce to that of averages $\mean{\mathcal{J}_\ell}$ and correlation function $\G(\mathcal{J}_\ell, \mathcal{J}_{\ell'})$ of the spin operator.

We first consider LS coherent states $\ket{LS,\{\lambda_{n\mu},\vec{v}_{n\mu}\}}$, defined in Eq.~\eqref{eq:def-coherent-states}, with the choice of parameters $\{\lambda_{n\mu},\vec{v}_{n\mu}\}$ specified in Eqs.~\eqref{eq:cubic-pars-1} and \eqref{eq:cubic-pars-2} on a cubic lattice. In the limit of large spins, we have shown that the system decomposes into a collection of independent one-dimensional systems.  Consider the one-dimensional lattice $\Gamma_{x00}=\{(n_1,0,0) \mid n_1=1,\ldots,L\}$. The fluctuations of the spins in this sublattice are independent from those of spins in its complement. Denoting the spin at the link to the right of the vertex $(p,0,0)$ by
\begin{equation}	\label{eq:j-to-d}
 j_p=j_0+d_p \, ,
 \end{equation}
the probability of a given configuration $\{j_p \}$ is described by the Gaussian probability distribution \eqref{eq:spin-prob-coherent}, with $\Delta_{na}$ defined in Eq. \eqref{eq:delta-def}, restricted to the independent sublattice $\Gamma_{x00}$:
\begin{equation}
P_\lambda(\{j_p\}) \propto \exp\left[ -\frac{1}{2} \sum_{rs} A_{rs} d_r d_s \right] \, ,
\label{eq:1d-gaussian}
\end{equation}
with:
\begin{equation}\label{eq:1d-gaussian-ab}
A = \frac{10}{\lambda^2} \left( \mathbb{I}- \frac{1}{10} B \right)\, , \quad B = \begin{bmatrix}
0	& 	1	&	0	&	\cdots	&	0	&	1\\
1	& 	0	&	1	&	0	&	&	0 \\
0	& 	1	&	0	&	1	&	0	&	\vdots \\
\vdots	& 	0		&	1	&	0	&	1	&	0\\
0	& 		&	0	&	1	&	0	&	1 \\
1	& 	0	&	\cdots	&	0	&	1	&	0
\end{bmatrix} \, .
\end{equation}

Let us denote by $\mathcal{J}_r$ the spin operator of the link to the right of the vertex $(r,0,0)$ in $\Gamma_{x00}$. We wish to compute the correlation function $\G(\mathcal{J}_r, \mathcal{J}_s)$, which requires the calculation of the averages $\mean{\mathcal{J}_r}$ and $\mean{\mathcal{J}_r \mathcal{J}_s}$. Such expectation values can be computed from the probability distribution \eqref{eq:1d-gaussian} as classical averages of the variables $j_r$ and $j_ r j_s$, interpreted as classical variables. Then, from Eq.~\eqref{eq:j-to-d},
\begin{align}
\mean{\mathcal{J}_r} &= \mean{j_r} = j_0 \nonumber \\
\mean{\mathcal{J}_r \mathcal{J}_s} &= \mean{j_r j_s} = j_0^2 + \mean{d_r d_s} \, ,
\end{align}
and
\begin{equation}
\G(\mathcal{J}_r, \mathcal{J}_s) = \mean{d_r d_s} \, .
\end{equation}
Now, for a Gaussian probability distribution expressed in the form \eqref{eq:1d-gaussian}, we have:
\begin{equation}
\mean{d_r d_s}= [A^{-1}]_{rs} \, .
\label{eq:jj-inv-A}
\end{equation}

Since the system is translationally invariant, the correlations depend only on the lattice distance between $r$ and $s$, and one can focus on the calculation of:
\begin{equation}
\mathcal{G}(R) = \mean{d_1 d_{1+R}} \, .
\end{equation}
For lattice distances much smaller than the total length $L$ of the lattice, the boundary conditions should not influence the correlations, which allows one to neglect the components with numerical values at the corners of the matrix $B$. Moreover, the matrix $A$ can be inverted by using the Neumann series:
\begin{equation}
A^{-1} = \frac{\lambda^2}{10} \sum_n \left(\frac{B}{10}\right)^n \, .
\label{eq:inv-A}
\end{equation}
Because of the special form of $B$, having nonzero components only in the secondary diagonals above and below the main diagonal, it turns out that the main contribution for the matrix component $[A^{-1}]_{1(1+R)}$ is of $R$-th order, and reads:
\begin{equation}
\left[\sum B^n \right]_{1(1+R)} \simeq 1 \, .
\label{eq:A-inverse-approx}
\end{equation}
This leads to:
\begin{align}
\mathcal{G}(R) & \simeq \lambda^2 10^{-R-1} \notag \\
	& \simeq \frac{j_0}{5} e^{-2.3 R} \, , 
\label{eq:corr-length}
\end{align}
from which we read off the correlation length $\xi \simeq 0.43$. Therefore, we find that the correlations are short ranged, reaching only a few links in the lattice.

Let us now consider the case of heat kernel states $\ket{HK,\{H_\ell,t_\ell\}}$, defined in Eq.~\eqref{eq:def-coherent-states}, with the parameters $H_\ell$ decomposed as specified in Eq.~\eqref{eq:sl2c-decomp}, normal vectors given by Eq.~\eqref{eq:cubic-pars-1}, and parameters $\xi_\ell, \eta_\ell, t_\ell$ specified by Eq.~\eqref{eq:hs-pars-2}, so that the states are peaked on cubic lattices. Repeating the steps followed for the case of LS coherent states, we now find:
\begin{equation}
	A = \left(\frac{1}{j_0}+4t \right) \left( \mathbb{I}- \frac{1}{2(1+4tj_0)} B \right) \,,
\end{equation}
and the dominant contribution to the correlations is:
\begin{equation}
\mathcal{G}(R) \simeq \frac{j_0}{(1+4tj_0)} [2(1+4tj_0)]^{-R}
\end{equation}
The correlation length is:
\begin{equation}
	\xi = \frac{1}{\log[2(1+4tj_0)]} < \frac{1}{\log 2} \simeq 1.44 \, .
\end{equation}
The correlations are again short ranged, reaching only a few links in the lattice.

\section{Perturbed coherent states: Nonlocal correlations}
\label{sec:correlated-coherent-states}

We now introduce a family of states with prescribed long range correlations built as a deformation of LS coherent states. We have seen in the Section \ref{sec:coherent-states} that the correlations in the fluctuations of the geometry at distinct nodes decay exponentially over the lattice for coherent states. We will show that long-range correlations can be introduced as perturbations obtained by composing the spins of the coherent states with those of two-loop states
\begin{equation}
\ket{\alpha_1 \alpha_2} = \frac{1}{2^4} F^\dagger_{\alpha_1}F^\dagger_{\alpha_2} \ket{0} \, ,
\end{equation}
where $\alpha_1$ and $\alpha_2$ are disjoint elementary loops formed by four links, or plaquettes. We first define such an operation of spin composition in the bosonic representation, and then introduce the perturbed coherent states and determine their main properties. 

At each seed $i$,  the operation of addition of angular momentum can be represented by the unitary map
\begin{equation}
	T_i: \Hh_i \otimes \Hh_i \to \Hh_i
\end{equation}
defined by
\begin{multline}
	T_i \left( \ket{j_a m_a}_i \otimes \ket{j_b m_b}_i \right) \\
		= \sum_{J=\mid j_a - j_b \mid}^{j_a + j_b} \sum_{M=-J}^J C^{JM}_{j_a m_a j_b m_b} \ket{JM}_i \, ,
\end{multline}
where
\begin{equation}
	C^{JM}_{j_a m_a j_b m_b} = \scalar{Jm}{j_a m_a; j_b m_b} \, .
\end{equation}
are Clebsch-Gordan coefficients. Taking the tensor product over all seeds, we obtain a unitary map:
\begin{equation}
	T: \Hh_\mathcal{S} \otimes \Hh_\mathcal{S} \to \Hh_\mathcal{S}, \quad T = \bigotimes_{i=1}^{2L} T_i \, . 
\end{equation}
The map $T$ is well defined on the subspace of gauge-invariant states $P_G \Hh_\mathcal{S}$, as it preserves gauge-invariance. In order for the spins at the source and target nodes at each link to match, the projector $P_A$ must be afterwards applied. This procedure defines an operation that we denote by
\begin{equation}\label{eq:prod-h-def}
	\ket{\Psi_1} \cdot \ket{\Psi_2} = P_A T(\ket{\Psi_1} \otimes \ket{\Psi_2}) \, ,
\end{equation}
which is well-defined on the kinematical subspace $\K_\Gamma$. For brevity, we will say that the operation \eqref{eq:prod-h-def} is the tensor product of the states $\ket{\Psi_1}$ and $\ket{\Psi_2}$.

The spin composition defined by Eq.~\eqref{eq:prod-h-def} has a formal similarity to the pointwise product of wavefunctions in the holonomy representation,
\begin{equation}	\label{eq:pointwise-prod}
\psi_1(h_\ell) \psi_2(h_\ell) \, ,
\end{equation}
but let us note that the operations do not precisely agree. Consider, for instance, states of the form
\begin{equation}	\label{eq:D-times-D}
\prod_{\ell \in \Gamma} \sqrt{2j_\ell + 1} [D^{j_\ell}(h_\ell)]^{m_\ell}{}_{n_\ell} \, ,
\end{equation}
consisting of a product of coefficients of Wigner matrices over all links of the graph, which provide an orthonormal basis of $\Hh_\Gamma$, from Eq.~\eqref{eq:spin-h-map}. At each link, the product of two such basis vectors leads to an expression of the form:
\begin{multline}	\label{eq:CG-series}
\sqrt{2j_1 + 1} [D^{j_1}(h)]^{m_1}{}_{n_1} \sqrt{2j_2 + 1} [D^{j_2}(h)]^{m_2}{}_{n_2}  	 \, ,
\end{multline}
which reduces, from the Clebsch-Gordan series \cite{sakurai}, 
\begin{multline}	\label{eq:CG-series}
[D^{j_1}(h)]^{m_1}{}_{n_1} [D^{j_2}(h)]^{m_2}{}_{n_2} \\
	= \sum_{JMN} C^{JM}_{j_1 m_1 j_2 m_2} C^{JN}_{j_1 n_1 j_2 n_2} [D^{J}(h)]^{M}{}_{N} \, ,
\end{multline}
and the map \eqref{eq:spin-h-map}, to the form
\begin{multline}	\label{eq:prod-2}
\sum_{JMN} (-1)^{J-N} \frac{\sqrt{2j_1 + 1} \sqrt{2j_2 + 1}}{\sqrt{2J+1}} \\
	\times C^{JM}_{j_1 m_1 j_2 m_2} C^{JN}_{j_1 n_1 j_2 n_2} \ket{JM} \ket{J,-N}
\end{multline}
in the spin representation. On the other hand, the operation \eqref{eq:prod-h-def} produces a new link state of the simpler form
\begin{equation}	\label{eq:prod-1}
\sum_{JMN} (-1)^{J-N} C^{JM}_{j_1 m_1 j_2 m_2} C^{JN}_{j_1 n_1 j_2 n_2} \ket{JM} \ket{J,-N} \, ,
\end{equation}
without the factors involving square roots.

When applied to a generic state $\ket{\psi_0} \in \K_\Gamma$ and a two-loop state $\ket{\alpha_1 \alpha_2}$, the operation \eqref{eq:prod-h-def} corresponds to first acting on the vacuum state with the product of operators describing the traces of the holonomies along the loops and then composing the spins of the resulting two-loop state with those of $\ket{\psi_0}$. On the other hand, the pointwise product \eqref{eq:pointwise-prod} corresponds to directly acting with the same operator on the state $\ket{\psi_0}$. The more compact form of the composition of spins given by Eq.~\eqref{eq:prod-1} in comparison to that given by Eq.~\eqref{eq:prod-2} will be convenient for the definition of perturbed states that are more manageable for the calculation of spin correlations.

Let us now discuss how perturbations encoding long-ranged correlations can be introduced by exploring composition of spins of an unperturbed state $\ket{\psi_0}$ and of two loop states $\ket{\alpha_1 \alpha_2}$ through the application of the tensor product \eqref{eq:prod-h-def}. Before introducing the perturbed LS coherent states, let us first discuss the simpler case of a spin-network state $\ket{\Gamma, \{j_\ell, i_n\}}$. Such a state factorizes over the nodes of the graph, \ie, it displays independent fluctuations of the geometry at distinct nodes. Nonlocal correlations can be introduced as follows. By taking the tensor product of the spin-network state with a two-loop state $\ket{\alpha_1 \alpha_2}$, where $\alpha_1$ and $\alpha_2$ are disjoint elementary loops formed by four links, or plaquettes, we obtain a new state
\begin{equation}\label{eq:two-loop-correlated-def}
\ket{\Gamma, \{j_\ell, i_n\}, \alpha_1 \alpha_2} \propto \ket{\Gamma, \{j_\ell, i_n\}} \cdot \ket{\alpha_1 \alpha_2} \in \mathcal{K}_\Gamma \, .
\end{equation}
As the tensor product only affects spins along the loops, which can change as $j_\ell \to j_\ell \pm 1/2$, it follows that
\begin{multline}\label{eq:two-loop-correlated}
\ket{\Gamma, \{j_\ell, i_n\}, \alpha_{12}} \in \left(\bigotimes_{\ell \notin 
\alpha_1, \alpha_2} \Hh_{j_\ell} \otimes \Hh_{j_\ell}\right) \\
	\otimes \left( \bigotimes_{\ell \in \alpha_1, \alpha_2} \bigoplus_{j_\ell \pm \frac{1}{2}}  \Hh_{j_\ell} \otimes \Hh_{j_\ell} \right) \, .
\end{multline}
The state $\ket{\Gamma, \{j_\ell, i_n\}, \alpha_1 \alpha_2}$ includes correlations along the loops $\alpha_1, \alpha_2$. A state with correlations distributed over the whole graph can then be constructed by summing over such states for arbitrary pairs of plaquettes.

Let us now consider the case of semiclassical states. We have seen in the Section \ref{sec:coherent-states} that, for both LS coherent states and heat kernel states, the correlations in the fluctuations of the geometry at distinct nodes decay exponentially over the lattice. Long-range correlations can be introduced as perturbations obtained by taking tensor products with two-loop states. Let us illustrate the procedure for the case of LS coherent states $\ket{LS,\{\lambda_{n\mu},\vec{v}_{n\mu}\}}$. We adopt the choice of parameters $\{\lambda_{n\mu},\vec{v}_{n\mu}\}$ specified in Eqs.~\eqref{eq:cubic-pars-1}, associated with a cubic lattice. For simplicity, we fix the spins to be the same at all links:
\begin{align}	\label{eq:homogeneous-LS-fixed-j}
\ket{\Psi_0} &\propto P_{ \{j_\ell=j_0\}} \ket{LS,\{\lambda_{n\mu}=\lambda,\vec{v}_{n\mu} = \vec v_\mu\}} \nonumber \\
	&\propto \bigotimes_{n = 1}^{N} \ket{ LS,\{j_{n \mu}=j_0, \vec{v}_{n\mu}=\vec v_\mu\} }_n \, ,
\end{align}
where $P_{ \{j_\ell=j_0\}}$ projects into the subspace with fixed spins $j_\ell=j_0$. Such a state factorizes with respect to the nodes. We now introduce the perturbed state:
\begin{equation}\label{eq:correlated-LS}
		\ket{\Psi} = \ket{\Psi_0} + \gamma \sum_{\alpha_1, \alpha_2} c_{\alpha_1 \alpha_2} \ket{\Box\Box} \, ,
\end{equation}
with perturbations $\ket{\Box\Box}$ defined by:
\begin{equation}	\label{eq:two-loop-excitation}
		\ket{\Box\Box} = \mathcal{N} \, \ket{\Psi_0} \cdot \ket{\alpha_1 \alpha_2} \, ,
\end{equation}
where $c_{\alpha_1\alpha_2}$ are free parameters, $\gamma \ll 1$ is a small perturbative parameter, the sum runs over all pairs of disjoint elementary loops $\alpha_1$ and $\alpha_2$, and $\mathcal{N}$ is a normalization constant. The nonperturbed state $\ket{\Psi_0}$ is a semiclassical homogeneous background, and the perturbations represent correlated quantum fluctuations of the geometry over the background geometry, with a decay encoded, as we will see, in the function $c_{\alpha_1\alpha_2}$ that describes the variation of the amplitude of the two-loop excitations. This setup is inspired by the representation of the free graviton in the context of quantum field theory on curved spacetimes in terms of states characterized by a two-point function over a curved background.

Some properties of the state \eqref{eq:correlated-LS} can be analytically determined in the limit of large spins $j_0 \gg 1$ (see Appendix for details):
\begin{enumerate}
	\item Normalization: The nonlocal correlations introduce a correction on the norm of the state:
		\begin{equation}	\label{eq:norm-correlated-LS}
			\braket{\Psi}{\Psi} = 1 + \gamma^2 \sum_{\alpha_1 \alpha_2} \left(c_{\alpha_1 \alpha_2}\right)^2 \, .
		\end{equation}
	\item Spin probabilities: The probability distribution for the spins at each link $\ell$ acquires spin fluctuations around the background value $j_0$. Taking a partial trace on degrees of freedom in the region complementary to the link, we find:
	\begin{widetext}
		\begin{equation}	\label{eq:spin-distribution-correlated-LS}
			P(j_\ell) = \delta_{j_\ell, j_0} + \gamma^2  \sum_{ \alpha_1 \cup \alpha_2 \ni \ell} \left( c_{\alpha_1 \alpha_2}\right)^2 \left[ \frac{\delta_{j_\ell,  j_0+1/2}}{2} \left(1 + \frac{1}{j_0}\right) + \frac{\delta_{j_\ell,  j_0-1/2}}{2} \left(1 - \frac{1}{j_0}\right) - \delta_{j_\ell, j_0} \right] \, ,
		\end{equation}
	\end{widetext}
where the sum runs over all pairs of disjoint loops which include the link $\ell$.
	\item Average and dispersion of the spin at each link: From the probability distribution \eqref{eq:spin-distribution-correlated-LS}, we obtain:
	\begin{align}
			\mean{\mathcal{J}_\ell} &= j_0 +  \frac{\gamma^2}{2j_0} \sum_{\alpha_1 \cup \alpha_2 \ni \ell} \left( c_{\alpha_1 \alpha_2}\right)^2  \, , \\
            \mean{\mathcal{J}_\ell^2} &= j_0^2  + \frac{5\gamma^2}{4}   \sum_{\alpha_1 \cup \alpha_2 \ni \ell} \left( c_{\alpha_1 \alpha_2}\right)^2 
	\end{align}
leading to a relative fluctuation
 \begin{equation}\label{eq:fluctuations}
		\frac{\sigma_{j_\ell}}{\mean{\mathcal{J}_\ell}} =  \frac{\gamma}{2j_0} \sqrt{\sum_{\alpha_1 \cup \alpha_2 \ni \ell} \left( c_{\alpha_1 \alpha_2}\right)^2}
 \end{equation}
and the ratio vanishes in the limit of large spins.
\end{enumerate}

We see that the average value of the spins is not significantly affected by the perturbations: the spin fluctuations introduced by the perturbations are approximately symmetric around the mean value $\mean{\mathcal{J}_\ell} \simeq j_0$. The dispersion $\sigma_{j_\ell}$ remains small with respect to the mean value, and vanishes in the limit of $j_0 \to \infty$. The perturbed state thus remains semiclassical with respect to spins. In addition, long-range correlations are introduced by the perturbations, as we wish to discuss now.

Let us compute the correlations of area fluctuations for the state \eqref{eq:correlated-LS}. As discussed before in Sec. \ref{sec:correlations decay}, for large spins, this amounts to computing the correlation function for the spin operator,
\begin{equation}	\label{eq:correlation-function-2}
	\G(\mathcal{J}_\ell, \mathcal{J}_{\ell'}) = \mean{\mathcal{J}_\ell \mathcal{J}_{\ell'}} - \mean{\mathcal{J}_{\ell}} \mean{\mathcal{J}_{\ell'}} \, .
\end{equation}
The term $\langle \mathcal{J}_\ell \mathcal{J}_{\ell'}\rangle$ takes the form:
\begin{multline}\label{eq:<JJ>}
        \mean{\mathcal{J}_\ell \mathcal{J}_{\ell'}} = j_0^2 + \frac{\gamma^2}{2} \left[ \sum_{\alpha_1 \cup \alpha_2 \ni \ell} \left( c_{\alpha_1 \alpha_2}\right)^2 + \sum_{\alpha_1 \cup \alpha_2 \ni \ell' } \left( c_{\alpha_1 \alpha_2}\right)^2    \right. \\
        \left. + \frac{1}{2j_0^2}\sum_{\substack{\alpha_1 \cup \alpha_2 \ni \ell\\ \alpha_1 \cup \alpha_2 \ni \ell'}} \left( c_{\alpha_1 \alpha_2}\right)^2 \right] \, .
\end{multline}

The first two sums are volume terms, as they scale with the size of the graph. In contrast, the third sum is independent of the graph size and remains finite in the limit of an infinite graph. It is straightforward to show that, for a finite lattice, the volume terms cancel out in the calculation of the correlation function \eqref{eq:correlation-function-2}, leaving only the third sum in the contribution \eqref{eq:<JJ>}. As a result, the correlation function remains finite and independent of the graph size, and assumes the following form in the limit of large spins:
\begin{equation}\label{eq:corr-function-J}
		\G(\A_\ell, \A_{\ell'}) = a_0^2 \frac{\gamma^2}{4 j_0^2} \sum_{\substack{\alpha_1 \ni \ell \\ \alpha_2 \ni \ell'} } (c_{\alpha_1\alpha_2})^2 \, .
\end{equation}

The function $c_{\alpha_1\alpha_2}$ can be chosen arbitrarily on a finite lattice. In order for the limit of large graphs to be well-defined, the sum over pairs of loops in Eq.~\eqref{eq:corr-function-J} must converge. But this is true for any graph, as the sum involves only a finite number of terms. As the background is homogeneous, it is natural to let $c_{\alpha_1\alpha_2}$ be a function of the distance between the loops. By taking it to decay with the inverse of the distance between the links $\ell$, $\ell'$, we see that the correlation function reproduces the typical decay of correlations with $1/d^2$ for massless fields in the continuum.

Starting from an LS coherent state at fixed spins, the perturbed state introduced in Eq.~\eqref{eq:correlated-LS} is a simple modification that can be seen as a first contribution in a more general approach for the construction of correlated states of the geometry. Higher-order contributions can be envisage with perturbations involving, for instance, simultaneous excitations of a larger number of loops distributed in more than two locations of the graph, or longer loops than the considered plaquettes, as well as multiple excitations at a loop. As shown by Eq.~\eqref{eq:corr-function-J}, the simplest choice of perturbations describing entangled excitations of pairs of plaquettes is already sufficient for the construction of states displaying correlations decaying over the lattice with the typical behavior expected for massless free gravitons in semiclassical gravity.

\section{Discussion}
\label{sec:conclusion}

A basic requirement for semiclassical states in general is that they must be peaked on a classical configuration. This sets a condition on average values for the observables in question, which must be related as in a classical configuration, as well as on the dispersion of the observables, which must be small. These conditions can be satisfied by many distinct states, however, as the relevant observables may have the same averages and dispersions, but distinct correlations and higher-order correlation functions. Additional conditions can be envisaged to further restrict the selection of semiclassical states. In this work, we were concerned with the behavior of correlations in the fluctuations of the gravitational field for semiclassical states in loop quantum gravity. Consistency with the regime described by semiclassical gravity requires states of the geometry to display entangled fluctuations with correlations decaying as the inverse of the distance squared. We focused on calculations of the two-point correlation function for the area operator on a cubulation for semiclassical states describing a regular cubic geometry.

We first considered the case of Livine-Speziale coherent states and heat kernel states as examples of intrinsic and extrinsic coherent states in LQG. We found that the area-area correlation function decays exponentially on the lattice for such states, with a correlation length of only a few sites. Next, we introduced a new family of states obtained as perturbations $\ket{\Psi}=\ket{\Psi_0}+\ket{\delta \Psi}$ of LS coherent states $\ket{\Psi_0}$, and analytically computed the dominant contribution to the area-area correlation function. This allowed the identification of states with correlations decaying as $1/d^2$ in the limit of large quantum numbers, as desired. The method explored to introduce perturbations over the LS coherent states was based on angular momentum recoupling theory. A spin-network basis state $\ket{\Gamma, \{j_\ell, i_n\}}$ can be mapped into a state of a collection of spin systems living at the endpoints of the links of a graph, with spins at the source and target nodes equal to that of the link, $j_{s(\ell)}=j_{t(\ell)}=j_\ell$. For an elementary excitation of the gravitational field on a single loop, all spins along the loop are in a spin-$1/2$ state, while spins not crossed by the loop are in a spin-$0$ state. By considering the addition of the spins of a two-loop state $\ket{\alpha_1 \alpha_2}$ to those of an LS coherent state $\ket{\Psi_0}$, as described in Eq.~\eqref{eq:two-loop-excitation}, a state $\ket{\Box \Box}$ is obtained with the same average areas and relative dispersions that remain small in the limit of large spins. Hence, the peakedness property of the unperturbed state is preserved. Considering perturbations $\ket{\delta \Psi}$ described by superpositions of such two-loop deformations of LS coherent states for arbitrary pairs of loops $\alpha_1, \alpha_2$, weighted by a function $c_{\alpha_1 \alpha_2}$, the resulting pertubed states given in Eq.~\eqref{eq:correlated-LS} proved well-suited for analytical calculations and included states with the desired properties.

States of the geometry with long-ranged correlations in loop quantum gravity were previously discussed in \cite{Bianchi:2016tmw}. The states introduced in this work differ from squeezed vacuum states in two main respects. First, instead of introducing correlated excitations over the Ashtekar-Lewandowski vacuum state \cite{AL-vacuum,AL-vacuum-Baez}, as done for squeezed vacua, we considered excitations over Livine-Speziale coherent states. The semiclassical geometry described by the coherent states can then be interpreted as a background geometry over which correlated perturbations are introduced, in a picture reminiscent of perturbative quantum gravity. In addition, we directly considered correlated gauge invariant loop excitations, instead of excitations each involving, in the bosonic representation of LQG, a pair of bosonic variables at distinct regions of the lattice. This simplifies the calculation of the correlations, allowing nonzero corrections to unperturbed results to be obtained at a lower perturbation order in comparison with the case of squeezed vacua. Moreover, the resulting states are automatically gauge invariant.

A natural extension of the method introduced for the construction of perturbed coherent states consists in exploring perturbations obtained by adding the angular momenta of states with a larger number of gauge invariant excitations on two loops or excited over a larger number of loops to those of the unperturbed states. This would allow the construction of perturbed states with a richer network of correlations. We analyzed here the case of elementary two loop excitations over LS coherent states, and more general perturbations can be considered. We also note that, as for the usual families of intrinsic and extrinsic coherent states, our perturbed states are defined in the kinematical Hilbert space $\K$ of LQG, formed by states that are invariant under gauge transformations and spatial diffeomorphism. Physical states of the geometry must also satisfy the Hamiltonian constraint $\mathcal{C}$. Considering the family of perturbed coherent states as an ansatz, for a given representation of the Hamiltonian constraint, one can pose the question of how the norms of states $\mathcal{C}^n \ket{\Psi}$ are affected by the presence of correlations, and whether they can be minimized by a specific choice of perturbation. The underlying motivation for the construction of states with long-ranged correlations was that of reproducing basic features of the gravitational field in the semiclassical regime, which hint at properties that semiclassical solutions of the Hamiltonian constraint should display in full quantum gravity. An analysis of the interplay between the presence of correlations and the action of the Hamiltonian constraint for semiclassical states provides a strategy for investigating whether the presence of correlations leads to better approximations to semiclassical solutions of the constraint or not, which can be pursued in future works.

\appendix*
\section{Reduced density matrix for a single link}

Equation \eqref{eq:correlated-LS} corresponds to the perturbed state on the entire lattice. To calculate expected values at a single link $\ell$, we first need to trace out degrees of freedom external to the link, and then compute the expected values for the resulting reduced density matrix. We describe the main steps required for that in this appendix.

We are interested in computing averages $\mean{\mathcal{J}_\ell}$ and $\mean{\mathcal{J}_{\ell}^2}$ for the perturbed coherent states $\ket{\Psi}$ in the limit of large spins. In this limit, for LS coherent intertwiners with parameters satisfying
\begin{equation}
\sum_\mu j_{n \mu}\vec{v}_{n \mu} = \vec{0} \, ,
\end{equation}
corresponding to the classical version of the Gauss constraint $\vec{G}_n$, the calculation of averages of gauge-invariant observables for the LS coherent states can be performed with the replacement
\begin{equation}	\label{eq:LS-large-j-approx}
\ket{ LS,\{j_{n \mu}, \vec v_{n \mu} \} }_n \to \bigotimes_{\mu=1}^{|n|} \ket{j_{n\mu},z_{n \mu}}_{n \mu} \, ,
\end{equation}
\ie, neglecting the action of the Gauss projector \cite{LSvertex}. We comment more on this approximation at the end of this appendix. Adopting it, the unperturbed state $\ket{\Psi_0}$ is a tensor product of $SU(2)$ coherent states of the form
\begin{equation}
\ket{j_0,\pm j_0}_a
\end{equation}
at individual seeds, where $\ket{j,m}_a$, $a=x,y,z$, are basis vectors in the magnetic number representation for spin projections along the axes $x,y,z$, matching the direction of the link associated with the seed in the lattice.

The perturbations $\ket{\square \square}$ are obtained by taking tensor products with two-loop states (Eq.~\eqref{eq:two-loop-excitation}). Only links crossed by the loops are affected. Consider, for concreteness, and without loss of generality, due to the symmetries of the setup, a link $\ell$ along the direction $z$ crossed by a loop coming from the direction $x$. The state at the union of the two wedges of the loop intersecting $\ell$ is, up to a numerical factor,
\begin{multline} \label{eq:link-state-perturbed}
T[ (\ket{j_0,j_0}_x \ket{j_0,j_0}_z \ket{j_0,-j_0}_z \ket{j_0,-j_0}_x \\
	\otimes \left( (\ket{-+}_z-\ket{+-}_z) \otimes (\ket{-+}_z-\ket{+-}_z) \right) ] \, ,
\end{multline}
where $\ket{-+}_z=\ket{1/2,-1/2}_z \otimes \ket{1/2,+1/2}_z$, etc, and the factors in the first and second line in \eqref{eq:link-state-perturbed} correspond to the restrictions of $\ket{\Psi_0}$ and $\ket{\alpha_1 \alpha_2}$ to the two-wedge region, respectively. The state at the link is described by the second and third factors of the tensor product of four factors in Eq.~\eqref{eq:link-state-perturbed}.

The contributions from each of the four terms resulting from (the label $z$ will be ommited in the kets from now on)
\begin{multline}
(\ket{-+}-\ket{+-}) \otimes (\ket{-+}-\ket{+-}) \\
	= \ket{-+}\ket{-+} - \ket{-+} \ket{+-} - \ket{+-}\ket{-+} + \ket{+-}\ket{+-}   
\end{multline}
are orthogonal. After the product with the loop state, the contribution of each of these four terms is a tensor product of seed states of the form:
\begin{align}
\ket{j_0,\pm j_0 ,\pm} &:= T_i(\ket{j_0,\pm j_0} \otimes \ket{\pm}) \nonumber \\
	&= \ket{j_0+\frac{1}{2}, \pm j_0 \pm \frac{1}{2}} \, ,\\
\ket{j_0,\pm j_0,\mp} &:= T_i(\ket{j_0,\pm j_0} \otimes \ket{\mp}) \nonumber \\
	&= \sqrt{\dfrac{1}{2j_0+1}}\ket{j_0+\frac{1}{2}, \pm j_0 \mp \frac{1}{2}} \nonumber \\  	& \quad \pm \sqrt{\dfrac{2j_0}{2j_0+1}}\ket{j_0-\frac{1}{2}, \pm j_0 \mp \frac{1}{2}} \, ,
\end{align}
and the state at the link, after area-matching, is a superposition of four orthogonal states, given by: 
\begin{widetext}
\begin{align}\label{eq:link-contributions-pert}
P_A T( \ket{j_0,j_0} \ket{j_0,-j_0} \otimes \ket{+-}) &= \ket{j_0+1/2,j_0+1/2} \ket{j_0+1/2,-j_0-1/2} \, ,\nonumber \\
P_A T( \ket{j_0,j_0} \ket{j_0,-j_0} \otimes \ket{++}) &= \frac{1}{\sqrt{2 j_0+1}} \ket{j_0+1/2,j_0+1/2} \ket{j_0+1/2,-j_0+1/2} \nonumber \\
P_A T( \ket{j_0,j_0} \ket{j_0,-j_0} \otimes \ket{--}) &= \frac{1}{\sqrt{2 j_0+1}} \ket{j_0+1/2,j_0-1/2} \ket{j_0+1/2,-j_0-1/2} \nonumber \\
P_A T( \ket{j_0,j_0} \ket{j_0,-j_0} \otimes \ket{-+}) &= \frac{1}{2 j_0+1}\ket{j_0+1/2,j_0-1/2} \ket{j_0+1/2,-j_0+1/2} \nonumber \\
	& \quad - \frac{2j_0}{2j_0+1} \ket{j_0-1/2,j_0-1/2} \ket{j_0-1/2,-j_0+1/2}
\end{align}
\end{widetext}
In the full state over the whole lattice, each of these link states is multiplied by a relative state in the complement of the link, all of which have the same norm. It follows that the density matrix $\Trb{(\Gamma - \ell)}{\ket{\square \square}\bra{\square \square}}$ at the link is the mixture of the density matrices of each of the four states in \eqref{eq:link-contributions-pert}.

For the full perturbed state $\ket{\Psi}$ defined in Eq.~\eqref{eq:correlated-LS}, given that the perturbations $\ket{\square \square}$ associated with distinct pairs of loops are orthogonal among themselves and with the unperturbed state, it follows that the reduced density matrix associated with a link $\ell$ is given by
\begin{multline}\label{estado pre traço}
	\rho_\ell = \Trb{(\Gamma - \ell)}{\ket{\Psi_0}\bra{\Psi_0}} \\
		+ \gamma^2 \sum_{\alpha_1, \alpha_2} \left( c_{\alpha_1 \alpha_2} \right)^2 \Trb{(\Gamma - \ell)}{\ket{\square \square}\bra{\square \square}} \,. 
\end{multline}
Computing it in the limit of large spins, $ j_0 \gg 1$, we find:
\begin{widetext}
\begin{multline}
	\rho_\ell = \left[ 1 + \gamma^2 \sum_{\ell \notin \alpha_1 \cup \alpha_2} \left(c_{\alpha_1 \alpha_2} \right)^2 \right] \ket{j_0,j_0,-j_0}\bra{\;} + \dfrac{\gamma^2}{2} \sum_{\ell \in \alpha_1 \cup \alpha_2} \left(c_{\alpha_1 \alpha_2} \right)^2 \left[ \ket{j_0 + \frac{1}{2}, j_0 + \frac{1}{2}, -j_0 - \frac{1}{2}} \bra{\;} \right. \\
	 + \dfrac{1}{2j_0} \ket{j_0 + \frac{1}{2}, j_0 + \frac{1}{2}, -j_0 + \frac{1}{2}} \bra{\;} +  \dfrac{1}{2j_0} \ket{j_0 + \frac{1}{2}, j_0- \frac{1}{2}, -j_0 - \frac{1}{2}} \bra{\;} \\
	 -\dfrac{1}{2j_0} \ket{j_0 + \frac{1}{2}, j_0 - \frac{1}{2}, -j_0 + \frac{1}{2}} \bra{j_0 - \frac{1}{2}, j_0 - \frac{1}{2}, -j_0 + \frac{1}{2} } \\
	 - \dfrac{1}{2j_0} \ket{j_0 - \frac{1}{2}, j_0 - \frac{1}{2}, -j_0 + \frac{1}{2} } \bra{j_0 + \frac{1}{2}, j_0 - \frac{1}{2}, -j_0 + \frac{1}{2}} \\
	\left. + \left(1 - \dfrac{1}{j_0} \right) \ket{j_0 - \frac{1}{2}, j_0 - \frac{1}{2}, -j_0 + \frac{1}{2}} \bra{\;} \right] \, ,
\end{multline}
\end{widetext}
where $\ket{X}\bra{ \ } := \ket{X}\bra{X}$ and $\ket{j,m,n}:= \ket{j,m} \ket{j,n}$. From the state above, the properties shown in \eqref{eq:norm-correlated-LS} to \eqref{eq:fluctuations} follow immediately.

Let us comment on the large spin approximation \eqref{eq:LS-large-j-approx}. The Gauss projector $P_G$ can be implemented through a group averaging procedure:
\begin{equation}
P_G \ket{\psi}= \frac{1}{|SU(2)|^N} \int_{[SU(2)]^N} dU_n U_g \ket{\psi} \, ,
\end{equation}
where the gauge transformations $U_g$ were defined in Eq.~\eqref{eq:gauge-transf}. As the spin operators $\mathcal{J}_\ell$ commute with gauge transformations $U_g$, any function of the spins commutes with the Gauss projector, $[f(\mathcal{J_\ell}), P_G]=0$. The unperturbed LS coherent state can be expressed as:
\begin{align}
\ket{\Psi_0} &= P_G \ket{\tilde{\Psi}_0} \, , \nonumber \\
\ket{\tilde{\Psi}_0} &= \bigotimes_{n = 1}^{N} \bigotimes_{\mu=1}^{|n|} \ket{j_0,z_\mu}_{n \mu} \, .
\end{align}
Hence,
\begin{align}	\label{eq:large-j-approx-av}
\frac{\bra{\Psi_0} f(\mathcal{J_\ell}) \ket{\Psi_0}}{\mathcal{N}^2} &= \frac{1}{\mathcal{N}^2} \bra{\tilde{\Psi}_0} P_G f(\mathcal{J_\ell}) P_G \ket{\tilde{\Psi}_0} \nonumber \\
	&= \frac{1}{\mathcal{N}^2} \bra{\tilde{\Psi}_0} f(\mathcal{J_\ell}) P_G \ket{\tilde{\Psi}_0} \nonumber \\
	&= \frac{1}{\mathcal{N}^2} \frac{1}{|SU(2)|^N} \int dU_n \bra{\tilde{\Psi}_0} f(\mathcal{J_\ell}) U_g \ket{\tilde{\Psi}_0} \, ,
\end{align}
where we used that $P_G^2=P_G$, and 
\begin{equation}
\mathcal{N} =  \left[ \frac{1}{|SU(2)|^N} \int dU_n \bra{\tilde{\Psi}_0} U_g \ket{\tilde{\Psi}_0} \right]^{1/2}
\end{equation}
is a normalization constant. Now, for large spins, the overlap between the $SU(2)$ coherent state $\ket{\tilde{\Psi}_0}$ and its rotated image $U_g \ket{\tilde{\Psi}_0}$ decreases fast for $U_g$ far from the identity. Assuming that, as a result, the matrix elements $\bra{\tilde{\Psi}_0} f(\mathcal{J_\ell}) U_g \ket{\tilde{\Psi}_0}$ are negligible for $SU(2)$ coherent states $\ket{\tilde{\Psi}_0}$ and $U_g \ket{\tilde{\Psi}_0}$ associated with significantly distinct classical configurations, \ie, for $U_n$ far from the identity, the integral in Eq.~\eqref{eq:large-j-approx-av} is dominated by contributions near the identity, and
\begin{equation}
\frac{\bra{\Psi_0} f(\mathcal{J_\ell}) \ket{\Psi_0}}{\scalar{\Psi_0}{\Psi_0}} \simeq \bra{\tilde{\Psi}_0} f(\mathcal{J_\ell}) \ket{\tilde{\Psi}_0} \, .
\end{equation}
The validity of this approximation was indeed established in \cite{LSvertex}. The average reduces to the classical value of the quantity on the configuration the state is peaked on.

Consider now the case of the perturbations $\ket{\square \square}$. Similarly (omitting the map $T$ in the tensor products, which commutes with gauge transformations),
\begin{widetext}
\begin{align}
\frac{\bra{\square \square} f(\mathcal{J_\ell}) \ket{\square \square}}{\mathcal{N}_{\square}^2} &= \frac{1}{\mathcal{N}_{\square}^2} (\bra{\alpha_1 \alpha_2} \otimes \bra{\tilde{\Psi}_0}P_G) P_A f(\mathcal{J_\ell}) P_A (\ket{\alpha_1 \alpha_2}\otimes P_G \ket{\tilde{\Psi}_0}) \nonumber \\
	&= \frac{1}{\mathcal{N}_{\square}^2} (\bra{\alpha_1 \alpha_2} \otimes \bra{\tilde{\Psi}_0})P_G P_A f(\mathcal{J_\ell}) P_A P_G (\ket{\alpha_1 \alpha_2}\otimes \ket{\tilde{\Psi}_0}) \nonumber  \\
	&= \frac{1}{\mathcal{N}_{\square}^2} (\bra{\alpha_1 \alpha_2} \otimes \bra{\tilde{\Psi}_0}) P_A f(\mathcal{J_\ell}) P_G P_A (\ket{\alpha_1 \alpha_2}\otimes \ket{\tilde{\Psi}_0}) \nonumber \\
	&= \frac{1}{\mathcal{N}_{\square}^2} \frac{1}{|SU(2)|^N} \int dU_n (\bra{\alpha_1 \alpha_2} \otimes \bra{\tilde{\Psi}_0}) P_A f(\mathcal{J_\ell}) P_A (\ket{\alpha_1 \alpha_2}\otimes U_g \ket{\tilde{\Psi}_0}) \, ,
\end{align}
\end{widetext}
where we used the gauge invariance of the loop states, $U_g \ket{\alpha_1 \alpha_2} = \ket{\alpha_1 \alpha_2}$, the fact that $[f(\mathcal{J_\ell}), P_G]=0$, $[P_A, U_g]=0$, $P_G^2= P_G$, and
\begin{multline}
\mathcal{N}_{\square} = \left[ \frac{1}{|SU(2)|^N} \int dU_n (\bra{\alpha_1 \alpha_2} \otimes \bra{\tilde{\Psi}_0}) \right.P_A \\
	\left. (\ket{\alpha_1 \alpha_2}\otimes U_g \ket{\tilde{\Psi}_0}) \right]^{1/2}
\end{multline}
is a normalization constant. As for the unperturbed state, the overlap between the states $\ket{\alpha_1 \alpha_2}\otimes \ket{\tilde{\Psi}_0}$ and its rotated image $\ket{\alpha_1 \alpha_2}\otimes U_g \ket{\tilde{\Psi}_0}$ decreases fast for $U_g$ far from the identity, since it is equal to the overlap between $\ket{\tilde{\Psi}_0}$ and $U_g \ket{\tilde{\Psi}_0}$. We assume that, as a result, the integration is dominated by contributions near the identity, so that
\begin{equation}
\frac{\bra{\square \square} f(\mathcal{J_\ell}) \ket{\square \square}}{\mathcal{N}_{\square}^2} \simeq (\bra{\alpha_1 \alpha_2} \cdot \bra{\tilde{\Psi}_0}) f(\mathcal{J_\ell}) (\ket{\alpha_1 \alpha_2}\cdot \ket{\tilde{\Psi}_0}) \, .
\end{equation}
Under this approximation, for the calculation of averages of operators $f(\mathcal{J_\ell})$, the contribution of the perturbations to the link density matrix $\rho_\ell$ in Eq.~\eqref{estado pre traço} can be computed without applying the Gauss projector in the construction of the unperturbed coherent states. In particular, the probability distribution for the spins, $P(j_\ell)$, can be computed without applying the Gauss projector.

\end{document}